\documentclass[]{pasj02} 
\usepackage[switch,mathlines]{lineno} % add line number to manuscript

\jyear{2024}
\Received{}%{yyyy/mm/dd}
\Accepted{}%{yyyy/mm/dd}
\graphicspath{{./}{figures/}}

\newcommand{{\maxi}}{{\it MAXI}}
\newcommand{{\swift}}{{\it Swift}}
\newcommand{\srcname}{{MAXI J0709$-$159}}

\usepackage{url}
\usepackage{color}
\usepackage{amsmath}
\usepackage{multirow}
\usepackage{threeparttable}
\usepackage{natbib}
\usepackage{tabularx}

\begin{document} 

\title{Optical Observations of the High Mass X-ray Binary MAXI J0709$-$159/LY CMa}

%%% begin:list of authors
% Do NOT capitalize all letters in "textsc".
\author{
 Megumi \textsc{Shidatsu},\altaffilmark{1}\altemailmark\orcid{0000-0001-8195-6546} 
 \email{shidatsu.megumi.wr@ehime-u.ac.jp}
 Nobuyuki \textsc{Kawai}\altaffilmark{2}\orcid{0000-0001-9656-0261}
 Hiroyuki \textsc{Maehara}\altaffilmark{3,4}\orcid{0000-0003-0332-0811}
 Emi \textsc{Goto}\altaffilmark{5}
Sota \textsc{Urabe}\altaffilmark{5}
Wataru \textsc{Iwakiri}\altaffilmark{6}\orcid{0000-0002-0207-9010}
Yohko \textsc{Tsuboi}\altaffilmark{5}\orcid{0000-0001-9943-0024}
Noboru \textsc{Nemoto}\altaffilmark{5}\orcid{}
Sakura \textsc{Nawa}\altaffilmark{5}\orcid{}
Mutsumi \textsc{Sugizaki}\altaffilmark{7,8}\orcid{0000-0002-1190-0720}
Motoki \textsc{Nakajima}\altaffilmark{9}\orcid{0000-0001-6183-3207}
Masafumi \textsc{Niwano}\altaffilmark{2}\orcid{0000-0003-3102-7452}
Ryohei \textsc{Hosokawa}\altaffilmark{2}
Marie \textsc{Sakamoto}\altaffilmark{1}
 and 
Yoshiki \textsc{Matsuoka}\altaffilmark{1}\orcid{0000-0001-5063-0340}
}

\altaffiltext{1}{Department of Physics, Ehime University, 2-5, Bunkyocho, Matsuyama, Ehime 790-8577, Japan}
\altaffiltext{2}{Department of Physics, Tokyo Institute of Technology, 2-12-1 Ookayama, Meguro-ku, Tokyo 152-8551, Japan}
\altaffiltext{3}{Okayama Observatory, Kyoto University, 3037-5 Honjo, Kamogatacho, Asakuchi, Okayama 719-0232, Japan}
\altaffiltext{4}{Subaru Telescope Okayama Branch Office, National Astronomical Observatory of Japan, National Institutes of Natural Sciences, 3037-5 Honjo, Kamogata, Asakuchi, Okayama 719-0232, Japan}
\altaffiltext{5}{Department of Physics, Faculty of Science and Engineering, Chuo University, 1-13-27 Kasuga, Bunkyo-ku, Tokyo 112-8551, Japan}
\altaffiltext{6}{International Center for Hadron Astrophysics, Chiba University, Chiba 263-8522, Japan}
\altaffiltext{7}{National Astronomical Observatories, Chinese Academy of Sciences, 20A Datun Rd, Beijing 100012, China}
\altaffiltext{8}{Advanced Research Center for Space Science and Technology, Kanazawa University, Kakuma, Kanazawa, Ishikawa 920-1192, Japan}
\altaffiltext{9}{School of Dentistry at Matsudo, Nihon University,  2-870-1 Sakaecho-nishi, Matsudo, Chiba 101-8308, Japan}

%\footnotetext[$\dag$]{Present address: ....}

%%% end:list of authors

%% !!! Select 3 to 5 words from PASJ's key words !!! 
%% List of Key Words: https://academic.oup.com/pasj/pages/Pasj_Keywords 
%% "\KeyWords{ }" always has to be placed before ``\maketitle'' 
\KeyWords{X-rays: binaries --- stars: individual (MAXI J0709$-$159/LY CMa) --- stars: emission-line, Be --- accretion, accretion disks}  

\maketitle

\begin{abstract}
We report on the optical spectroscopic monitoring of the X-ray transient MAXI J0709$-$159 (identified as the Be star LY CMa) performed for about 1.5 months after the X-ray detection with MAXI. The observed spectrum showed a double-peaked H$\alpha$ line with a peak-to-peak separation of $\sim 230$ km s$^{-1}$, suggestive of the Be disk origin. We also detected a broad wing of the H$\alpha$ line with a line-of-sight velocity of $\gtrsim 900$ km s$^{-1}$, which could be explained by the accretion disk of the compact object or a stellar wind from the Be star. Initially the H$\alpha$ line showed an asymmetric profile with an enhanced blue peak, and then the blue peak decreased in $\sim$ 3 weeks to a similar strength to the red peak. 
We suggest that the evolution of the blue peak is associated with the X-ray activity and generated by the turbulence of the Be disk due to the passage of the compact object. We also investigated flux variation using the archival TESS data and found quasi-periodic variations with frequencies of $\sim 1$ and $\sim 2$ day$^{-1}$, which were likely caused by the pulsation of the B star. The overall variability properties on timescales of $\sim$ day were similar to those in Be X-ray binaries, rather than supergiant X-ray binaries.  
\end{abstract}

%\pagewiselinenumbers 

\section{Introduction}  \label{sec:intro}
% 1 introduction

High mass X-ray binaries (HMXBs), consisting of a 
compact object (a neutron star in majority 
but a black hole in some cases) and a massive star 
with $\gtrsim 10 M_\odot$, have several subclasses 
with a variety of observed X-ray and optical 
behavior \citep[see][for reviews]{chaty2011,walter2015,kretschmar2019}. 
They produce X-ray emission via accretion onto 
the compact object and optical emission mainly 
at the surface of the massive companion star 
and its surrounding structures. 

Well-known subclasses of HMXBs include the Be X-ray 
binary (BeXRB), hosting a non-supergiant 
fast-rotating B star with a circumstellar 
disk which emits strong H emission lines \citep[e.g.,][]{raig2011}. 
In these systems, accretion is driven by the 
interaction between the disk and the compact 
object. Another popular subclass is a supergiant 
X-ray binary (sgXB), whose compact object accretes 
matter by capturing the strong 
stellar wind from the supergiant donor. Their 
optical spectra often exhibit emission lines, 
sometimes associated with blueshifted absorption 
lines (so-called p-Cygni profile) produced by the 
wind. Many sgXBs are persistent X-ray sources, 
but some of them, called supergiant fast X-ray 
transients (SFXTs), show transient behavior. 
They are usually faint (typically at $\sim 10^{32}$ 
erg s$^{-1}$), but sporadically increase 
its X-ray luminosity by several orders of magnitude on 
timescales of $\sim$ hours \citep[e.g.,][]{walter2015}. 
The observed behaviors of HMXBs, including long-term 
optical and X-ray flux variability (such as the frequency 
and duration of the outbursts in transient sources), 
optical emission and absorption line profiles, 
and its variation during the outbursts, depend on the properties 
of the compact object, the donor star and its surrounding 
structures, and the binary orbit. Studies of HMXBs 
therefore enables us to investigate accretion processes 
and circumstellar environment of the massive donor stars, 
which are key information for binary evolution and perhaps
for their associations/non-associations with gravitational 
wave sources. 

MAXI J0709$-$159 is a transient Galactic X-ray binary discovered 
on 2022 January 25 \citep{serino2022ATel}, 
with Monitor of All-sky X-ray Image \citep[MAXI;][]{matsuoka09} via 
its nova alert system \citep{negoro16}. 
Follow-up X-ray observations were carried out 
\citep[][Sugizaki et al. in prep]{iwakiri2022ATel,
kobayashi2022ATel,negoro2022ATel, sugizaki2022} using MAXI, 
the Neutron star Interior Composition ExploreR 
\citep[NICER;][]{gendreau2012}, and the Nuclear 
Spectroscopic Telescope Array \citep[NuSTAR;][]{harrison2013}. 
From the refined source positions obtained with NICER and NuSTAR, 
the optical counterpart was identified as the Be star LY CMa/HD 54786, located at $\sim$ 3 kpc \citep{kobayashi2022ATel, sugizaki2022}. The source showed transient X-ray behavior 
consistent with 
SFXTs; it showed a rapid flux decay by $\sim$ 5 orders of magnitude in 3.5 days, 
associated with a strong spectral variation on timescales of $\sim$ 
hour with a change in the absorption column density by 1 order of 
magnitude \citep{sugizaki2022}. 
\citet{bhattacharyya2022} suggested the nature of the optical 
counterpart is an evolved B star, using optical and near-infrared 
photometric data. They also detected the H$\alpha$ emission line through 
optical spectroscopy $\sim$ 1 week after the first X-ray detection, indicating the presence of a Be disk. Comparing the result with those from previous spectroscopic observations, they found that the observed H and He emission lines were enhanced near the X-ray active period, suggesting possibilities that the circumstellar environment of the B star was affected by the passage of the compact object and/or that the Be disk increased its size and caused the X-ray activity. These observed properties suggest that MAXI J0709$-$159 is a somewhat unusual HMXB having both characteristics of BeXRBs and SFXTs. A recent report of further optical and near-infrared studies suggests that the source may be an evolutionary phase from BeXRB towards sgXB \citep{bhattacharyya2024}.

To further investigate the evolution of H$\alpha$ and other emission lines after the X-ray activity, we performed optical spectroscopy of the source from the end of January to the middle of March. We also investigated optical flux variability using archival data of Transiting Exoplanet Survey Satellite \citep[TESS;][]{ricker15}. 
In this article, we report these results and describe the possible origins of the observed emission lines and flux variation. 
We discuss the properties of circumstellar structure of 
the high-mass star and the accretion processes onto the compact 
object in this binary system.

%observation & data reduction 
\section{Observation and Data Reduction}
\label{sec:obs}

\begin{table*}
\begin{threeparttable}
  \centering
    \tbl{Observation log.}{
    \begin{tabularx}{125mm}{ccc}
  \hline \hline
  {\bf Observatory (Instrument)} & {\bf Obs. Date \& Start--End Time (UT)} & {\bf Exposure (s)}%\tnote{$\dagger$} 
  \\ \hline
      SCAT\tnote{$\dagger$} & 2022 Jan. 28 11:32-14:48 & 8340 (60 s $\times$ 139 frames) \\
   & 2022 Jan. 29 11:15-14:32 & 8460 (60 s $\times$ 141 frames) \\
   & 2022 Jan. 31 11:01-14:48 & 12960 (120 s $\times$ 108 frames) \\
   & 2022 Feb. 1 12:49-14:48 & 6840 (120 s $\times$ 57 frames) \\
   & 2022 Feb. 2 11:33-13:49 & 7800 (120 s $\times$ 65 frames) \\
   & 2022 Feb. 4 11:08-14:35 & 11100 (60 s $\times$ 185 frames) \\
   & 2022 Feb. 5 11:27-14:32 & 9960 (60 s $\times$ 166 frames) \\
   & 2022 Feb. 6 13:04-14:37 & 5400 (120 s $\times$ 45 frames) \\
   & 2022 Feb. 7 11:42-13:17 & 5100 (60 s $\times$ 85 frames) \\
   & 2022 Feb. 12 11:35-13:53 & 7920 (120 s $\times$ 66 frames) \\
   & 2022 Feb. 14 11:44-12:32 & 2640 (60 s $\times$ 44 frames) \\
   & 2022 Feb. 18 11:37-13:10 & 5040 (60 s $\times$ 84 frames) \\
   & 2022 Feb. 24 11:56-14:19 & 8160 (120 s $\times$ 68 frames) \\
   & 2022 Feb. 25 12:04-13:31 & 5040 (120 s $\times$ 42 frames) \\
   & 2022 Feb. 26 12:10-12:43 & 1800 (60 s $\times$ 30 frames) \\
   & 2022 Feb. 27 11:55-13:21 & 4680 (60 s $\times$ 78 frames) \\
   %%% ... 
   Seimei/KOOLS-IFU & 2022 Feb. 2 10:52-10:58 & 300 (60 s $\times$ 5 frames) \\
  OAO/HIDES & 2022 Feb. 4 12:51-13:21 & 1800\\
  & 2022 Feb. 8 13:04-13:34 & 1800 \\
   & 2022 Feb. 11 13:21-13:51 & 1800 \\
   & 2022 March 15 10:27-10:57 & 1200 \\
  \hline
  \end{tabularx}}
  \label{tab:obslog}
  \smallskip
  \begin{tablenotes}[normal]
   \footnotesize
   \small
   \item[$\dagger$] Equipped with a ATIK 460EX CCD camera and a Shelyak Alpy 600 spectrometer.%     
    \end{tablenotes}
\end{threeparttable}
\end{table*}

\subsection{Spectroscopy}
We carried out optical spectroscopic observations 
from 2022 January 28 to March 15 using three Japanese 
telescopes: Spectroscopic Chuo University Astronomical Telescope (SCAT), 
3.8 m Seimei telescope of Kyoto University, and the 1.88 m telescope 
at Okayama Astrophysical Observatory (OAO). Table~\ref{tab:obslog} 
gives the observation log.

\begin{figure*}
\begin{center}
     \includegraphics[width=160mm]{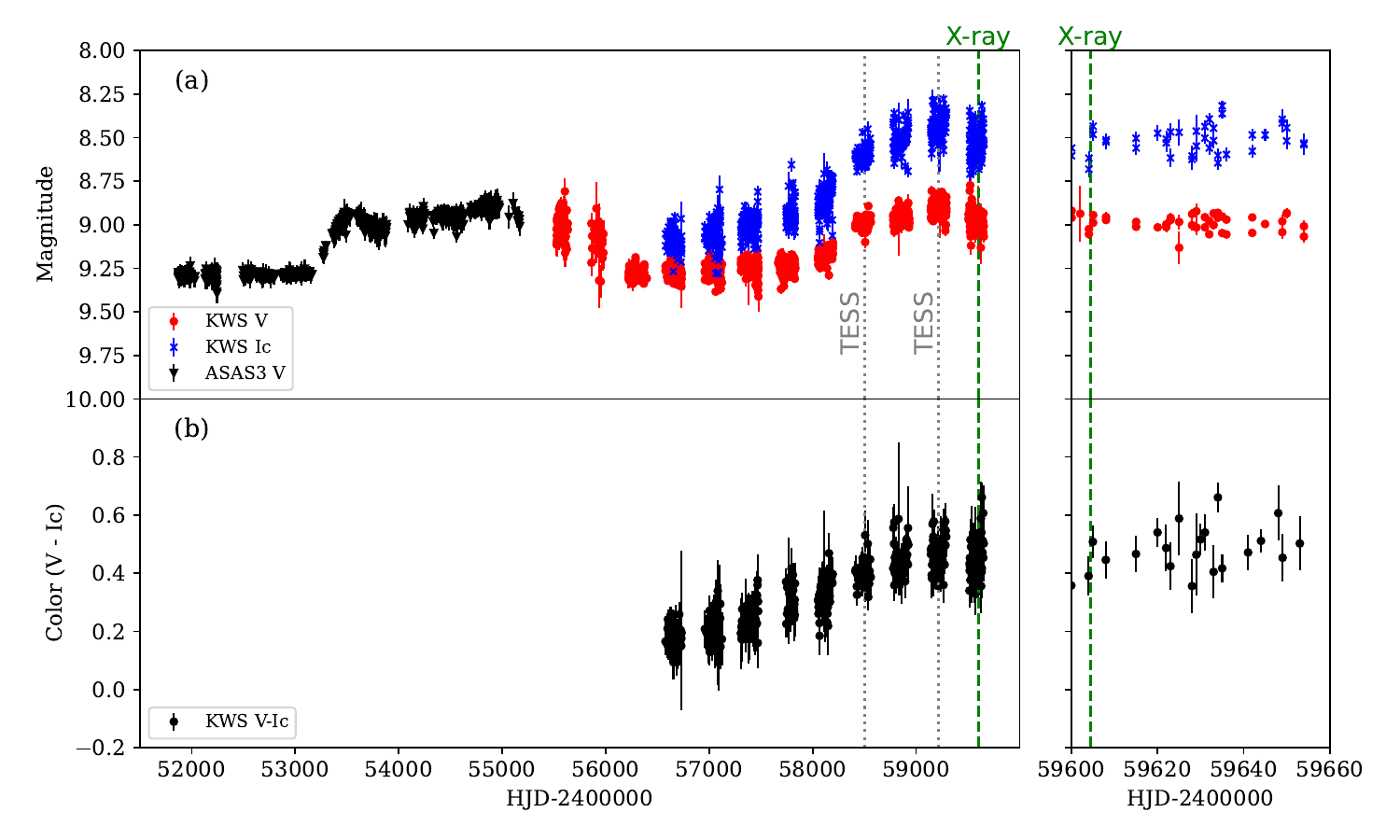}
 \end{center}
\caption{(a) long-term optical light curves of \srcname~produced from the ASAS3 V-band data (black triangles, until HJD $\sim 55000$) and KWS V-band and Ic-band data (red circles and blue crosses, respectively, after HJD $\sim 55000$) (b) variation of the V$-$Ic color (daily averages) obtained from the KWS data. The epochs of the TESS observations and the first X-ray detection are indicated. The right panels show enlarged views of the left panels, after the X-ray event. HJD 59600 corresponds to 2022 January 20. 
{Alt text: Graphs on the long-term variation of the V-band and Ic-band magnitudes and V$-$Ic color, with two vertically-separated panels labeled (a) and (b), which present the magnitude variations and color variation, respectively. A zoomed views of them in the range MJD 59600-–59660 are shown on the right.}
\label{fig:LC_longterm}}
\end{figure*}

Figure~\ref{fig:LC_longterm} presents the long-term optical light curves and V$-$Ic color of LY CMa, the optical counterpart of~\srcname, produced by collecting the data from the ASAS-3 catalog\footnote{\url{http://www.astrouw.edu.pl/asas/?page=aasc}} and the Kamogata/Kiso/Kyoto Wide-field Survey (KWS) database \citep{maehara14}. The source showed optical flux variation by $\sim 0.5$ mag in the V-band and $\sim 0.8$ mag in the Ic-band on timescales of $\sim 2000$ days and became redder as they became brighter. The X-ray flaring event occurred around the last optical flux peak. Our spectroscopic observation campaign started 3 days after the X-ray flare. The V and Ic band magnitudes remained almost constant during the campaign.

\subsubsection{SCAT}

\label{subsec:SCAT}
We performed spectral monitoring of MAXI J0709$-$159 with the SCAT, located at the Chuo Univeristy Korakuen campus, Tokyo, Japan \citep{kawai22}. The SCAT is a 355-mm diameter optical telescope equipped with an ATIK 460EX CCD camera and a Shelyak Alpy 600 spectrometer, with a resolution of R$\sim$410 around H$\alpha$ line at 6563~\AA~and a wavelength range of {3700--7500}~\AA~\citep{kawai22}. Note that the spectrometer is designed to achieve a resolution of R $\sim$ 600 \AA, but during this observing campaign, the resolution was somewhat degraded likely due to a slight shift of the focus  from the detector surface.

We reduced the SCAT data frame by frame following the standard data reduction procedures for optical spectra: dark frame subtraction, flat fielding, sky subtraction, spectral extraction and wavelength calibration. We performed all the steps with $python$\textquotesingle s astronomical packages. In the sky subtraction, we extracted the data of blank-sky regions in the object frames for the sky spectrum, and executed the wavelength calibration using the comparison lamp data (Ar and Ne). We found, however, that significant wavelength shifts were likely to remain even after the wavelength calibration; applying the same wavelength calibration to standard star data obtained at the same nights, we found that the absorption lines such as H$\alpha$ significantly shifted from the from the rest-frame wavelength. This could be because the comparison frames and the object frames were taken at different attitudes of the telescope and due to the the difference in the gravitational force affecting the optical system, the position of the spectral data on the image differed between the comparison frames and the object frames. 

We attempted to mitigate the systematic errors in the wavelength using the standard star data. We measured the line-center wavelengths of the H$\alpha$ lines in the standard star spectra at the individual nights and derived the offsets from the rest-frame value ($6562.8$~\AA). Then we shifted the wavelength of the entire object spectra by the offsets measured at the same nights. The typical offset was estimated to be 1--2~\AA. In some of the nights, however, we dit not conduct standard star observations, in which case we used the offset measured at the nearest night. We note that significant uncertainty of $\lesssim 1$~\AA~($\sim$ 2~\AA~in the case of no standard star data available) may remain even after this treatment.

Finally, we performed barycentric correction to the individual spectral data with the IRAF tool {\tt bcvcorr} %in the noao.rvsao package
 in the IRAF external package RVSAO \citep{kurtz1998}, although the barycentric velocity was estimated to be 6--21 km s$^{-1}$ (corresponding to 0.1--0.5~\AA~around the H$\alpha$ line), which was negligibly small compared with the wavelength resolution of the SCAT. We then obtained the final spectrum in each night by averaging the spectra obtained in the same nights. 

\begin{figure*}[t!]
  \includegraphics[width=160mm]{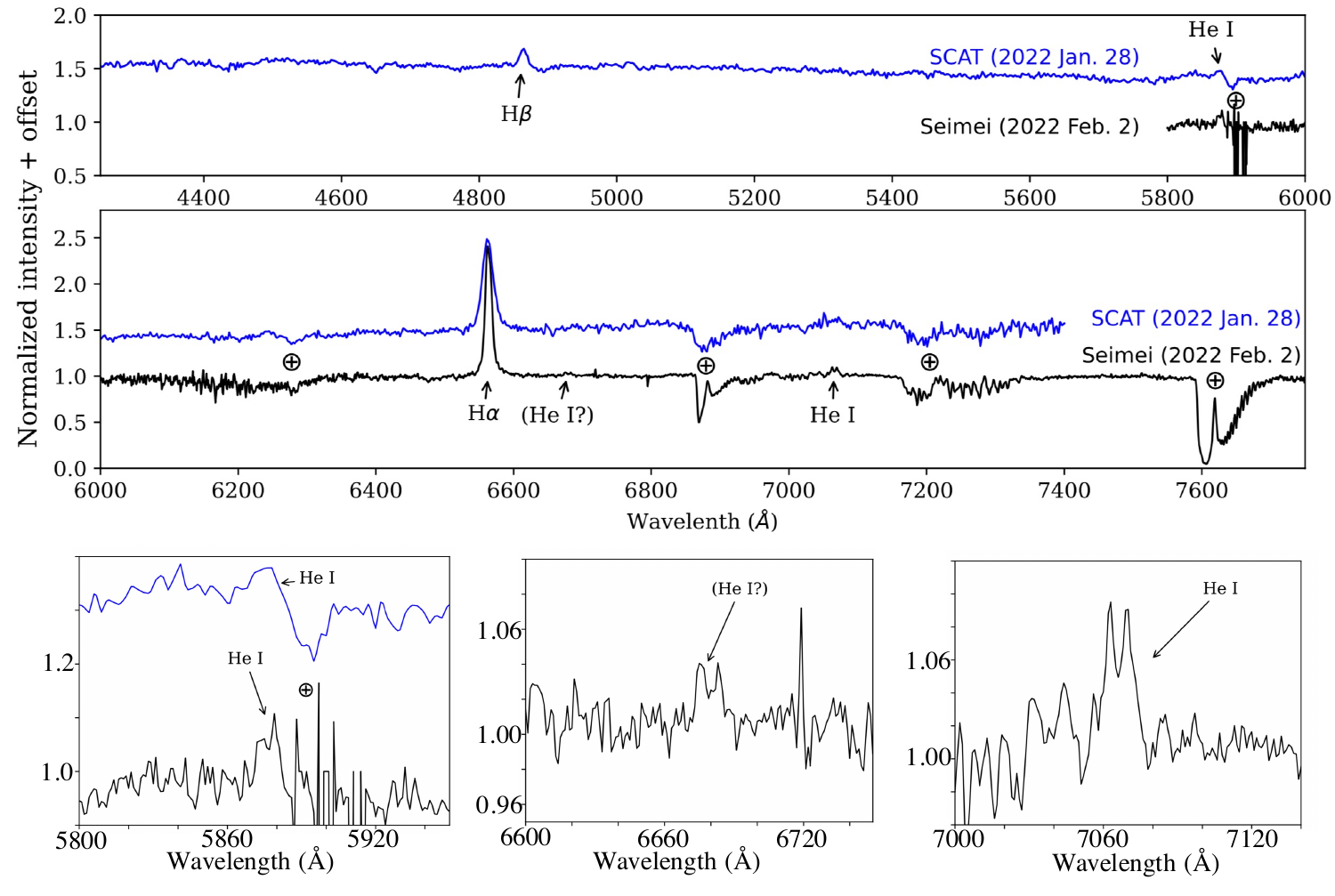}
\caption{Optical spectra of \srcname~obtained with the SCAT on 2022 January 28 (blue) and the Seimei/KOOLS-IFU on 2022 February 2 (black). 
Bottom panels show enlarged views around the weak HeI lines.
{Alt text: Spectral plots with three subfigures vertically. The top and middle subfigures show the spectra in 4200--6000~\AA~and 6000--7800~\AA. Bottom subfigure is divided into three panels horizontally, which are zoom-ins around HeI lines.}
}  
\label{fig:spec_full}
\end{figure*}

\subsubsection{Seimei}
\label{subsec:seimei}
On 2022 February 2, we performed a spectroscopic observation with the 3.8 m Seimei telescope of Kyoto University, located at Okayama Observatory of Kyoto University \citep{kurita10}, using the Kyoto Okayama Optical Low-dispersion Spectrograph with optical-fiber Integral Field Unit \citep[KOOLS-IFU][]{yoshida05,matsubayashi19}. We adopted the VPH-683 grism, which covers 5800--8000 \AA~with a wavelength resolution of $R=\lambda/\Delta \lambda \sim$ 2000. 

To reduce the Seimei data, we performed overscan and bias-pattern subtraction, spectral extraction, flat fielding, wavelength calibration, and then sky subtraction. 
The overscan, bias, and sky subtraction was performed with a python script for the KOOLS-IFU data reduction\footnote{http://www.kusastro.kyoto-u.ac.jp/\~kazuya/p-kools/reduction- 201806/install\_software.html}, while the other steps were with IRAF \citep{bar94, bar95}. The Hg and Ne comparison lamp data were used in the wavelength calibration. The sky spectrum was created from the object frames themselves, by collecting the data from fibers observing blank-sky regions. We conducted the above steps for the individual object frames and then took the median of their spectra. Barycentric 
correction was performed to the median spectrum with the IRAF tool {\tt bcvcorr} and used the resultant spectrum in the following analysis.

\subsubsection{HIDES} 
\label{subsec:HIDES}
We also carried out high-resolution optical spectroscopy at 4 nights in the period from 2022 February 4 to March 15, using Fiber-fed HIgh Dispersion Echelle Spectrograph \citep[HIDES-F;][]{karabe13} equipped on the 188 cm telescope at OAO. 
HIDES has two fiber optic link mode: HE (high efficiency) mode and HR (high resolution) mode, with wavelength resolutions 
of $R \sim 50000$ and $R \sim 100000$, respectively.
For our observations, we used the HE mode. 
The HIDES-F with the HE mode fiber link covers a wavelength range of 4100--7600 \AA. 
We reduced the OAO/HIDES data with the IRAF package {\tt echelle}, following the standard procedure for extracting spectra: overscan and bias subtraction, flat fielding, spectral extraction, and wavelength calibration based on Th-Ar lines. 
After obtaining a multi-order echelle spectrum, we performed blaze function correction, merging the spectral orders, continuum normalization, and barycentric wavelength correction.

\subsection{TESS observations}
To investigate short-term optical variability, we searched the archival TESS data and found that the source was observed twice: 2019 January 8--February 1 (hereafter we call Epoch-1) and 2020 December18--2021 January 13 (Epoch-2). 
We retrieved the light curve data at the two epochs from Mikulski Archive for Space Telescopes (MAST). 
The light curves, with 2-minute time bins, were 5$\sigma$ clipped to remove outlier data points. In the following analysis, we focused on the variability on timescales shorter than $\lesssim$ a few day. To remove variation on longer timescales, we detrended the light curves by calculating 3-day running average of the light curves and subtracting them from the original light curves.

% See the instraction below for "Alt text"
% https://academic.oup.com/pasj/pages/General_Instructions#Figures%20and%20Illustrations

\section{Analysis and Results}
\label{sec:ana}

\subsection{Overall Spectral Features}

\begin{figure}[thb]
\begin{center}
\includegraphics[width=90mm]{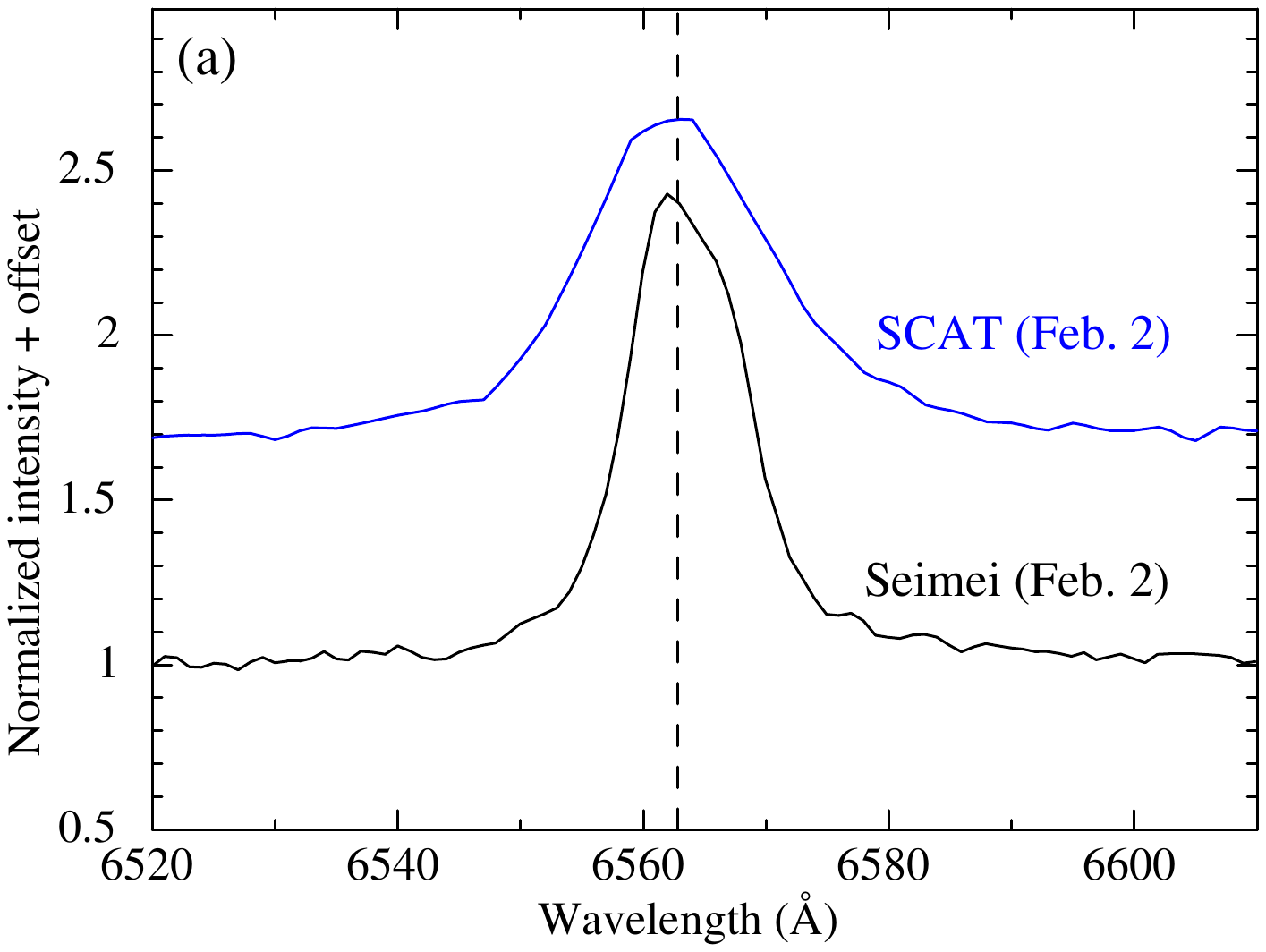}
\includegraphics[width=90mm]{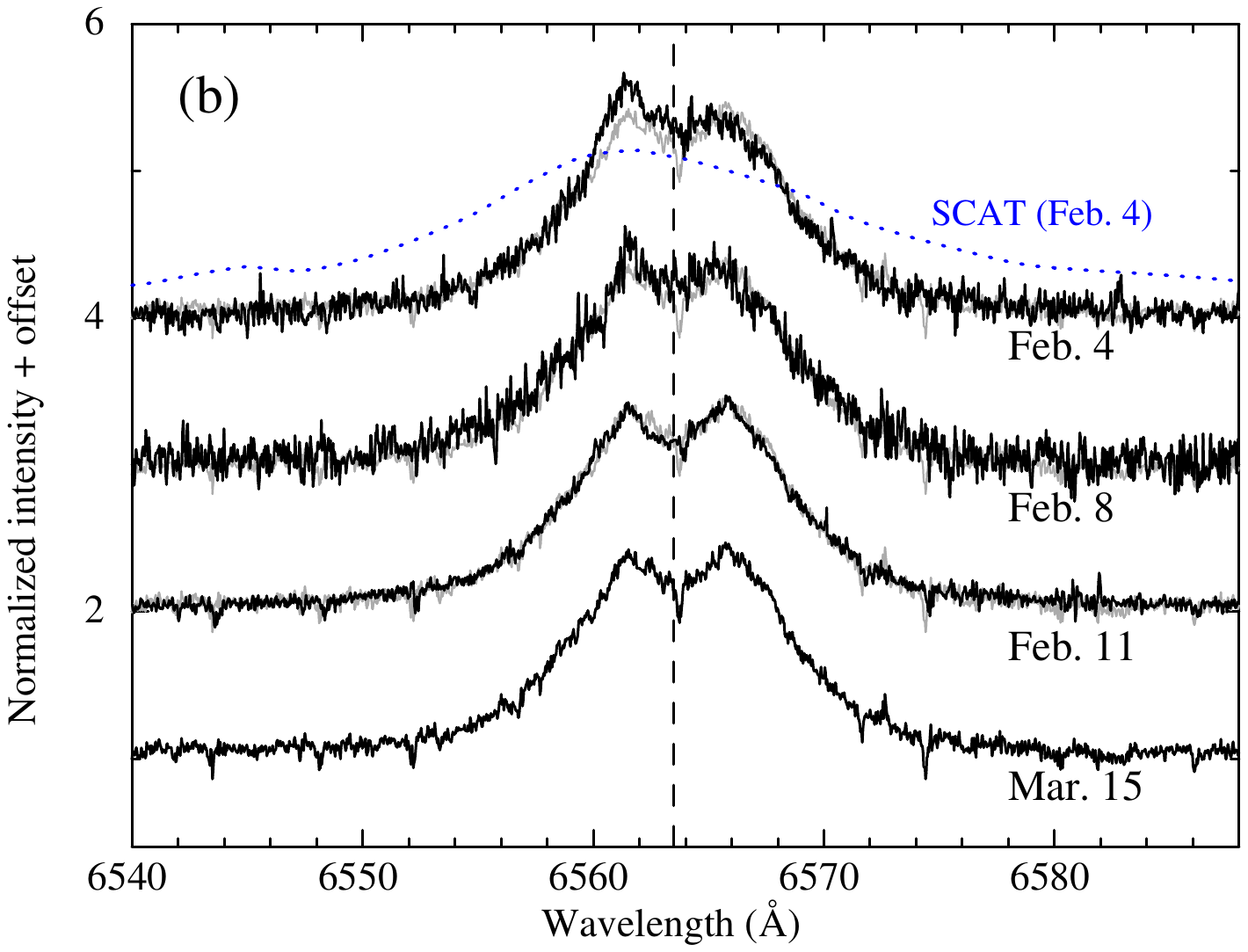}
 \end{center}
 \caption{H$\alpha$ line profiles in the Seimei/KOOLS-IFU and SCAT data on February 2 (a) and the profiles in the four OAO/HIDES spectra (b). Note that instrumental responses are included. The rest-frame wavelength of the H$\alpha$  lines are indicated with the dashed lines. The grey lines in the panel (b) show the Mar. 15 spectrum with the offset of the intensity adjusted to that of the other spectra for comparison. The SCAT spectrum on February 4 is also plotted to demonstrate the effects of the instrumental responses. 
{Alt text: Plots of the H$\alpha$ line profiles, with two vertically-aligned panels labeled (a) and (b) from top to bottom.}
\label{fig:halpha}}
\end{figure}

Figure~\ref{fig:spec_full} presents 
the SCAT and Seimei/KOOLS-IFU spectra obtained on Jan. 28 and Feb. 2, respectively, both of which show an H$\alpha$ emission line ($\lambda 6563$) clearly. We found that the H$\alpha$ was present in emission in the other SCAT spectra and that of the OAO/HIDES as well. 
In addition, we detected the H$\beta$ line ($\lambda 4861$) 
in the SCAT data and He I emission lines ($\lambda 5876$, $\lambda 7065$, and possibly $\lambda 6678$) in the SCAT and Seimei/KOOLS-IFU data. They were also detected in the OAO/HIDES spectra. A double-peaked profile was resolved clearly in these Balmer emission lines in the OAO/HIDES data (see Section~\ref{subsec:halpha} for the H$\alpha$ line). Some of the HeI lines in the OAO/HIDES and Seimei/KOOLS-IFU data also show hints of a double-peaked profile. HeII line ($\lambda 4686$) was not significantly detected in any of the SCAT data. 
 
In the following subsection, we focused on the H$\alpha$ line, which was the strongest feature in the spectra, and investigated the detailed line profile and its time evolution.

\subsection{Analysis of H$\alpha$ Emission Line} 
\label{subsec:halpha}

In Figure~\ref{fig:halpha}(a), we plotted an enlarged view around the H$\alpha$ line of the SCAT and Seimei/KOOLS-IFU spectra. The line in the Seimei spectrum has an asymmetric profile, with the bluer part enhanced. Such a profile was not clearly seen in the SCAT spectra, likely due to the limited spectral resolution. 
Figure~\ref{fig:halpha}(b) shows the H$\alpha$ line profiles 
obtained from high resolution spectroscopy with the OAO/HIDES. 
A double-peaked profile was clearly seen at all epochs. 
In the first two OAO/HIDES observation, the blue peak component 
was stronger than the red peak component, which is consistent with the asymmetry of the profile observed with Seimei/KOOLS-IFU. The line peak in the SCAT spectrum at the same night is slightly shifted to shorter wavelength and is consistent with the blue peak in the HIDES data, although it may be affected by the uncertainty in wavelength calibration (see Section~\ref{subsec:SCAT}). 
The blue peak decreased in later observations. Although much smaller than the variation of the blue peak, the red peak was also changed. The first two OAO/HIDES spectra showed slightly smaller red peaks than the latter two spectra. We found that 
the time variation was only seen around the two peaks (mainly the blue peak) and that the profile of both wings remained almost the same.

We analyzed the H$\alpha$ line profiles of the individual spectra with XSPEC version 12.12.1 included in HEASoft version 6.30.1. The spectral data in 6400--6700 \AA~were adopted and converted to the XSPEC format using the ftool {\tt ftflx2xsp} in the same 
HEASoft package. We estimated the standard deviation of the fluxes in the 
line-free regions around the H$\alpha$ line, 
6400--6500~\AA~and 6600--6700~\AA, for each spectra  
and adopted it as the flux error of each spectral bin. 
To consider their spectral resolution, we created a response 
matrix file for each instrument using the ftool 
{\tt ftgenrsp} such that the response is expressed as a Gaussian for each wavelength bin. We adopted full widths at half maximum (FWHMs) of 16 \AA, 3.4~\AA, and 0.12~\AA~for the SCAT, Seimei/KOOLS-IFU, and OAO/HIDES, respectively, which were estimated from the profile of an emission line near the H$\alpha$ line in the comparison lamp data. Hereafter the errors represent the 90\% confidence range for one parameter, unless otherwise stated.

\begin{figure*}
\includegraphics[width=60mm]{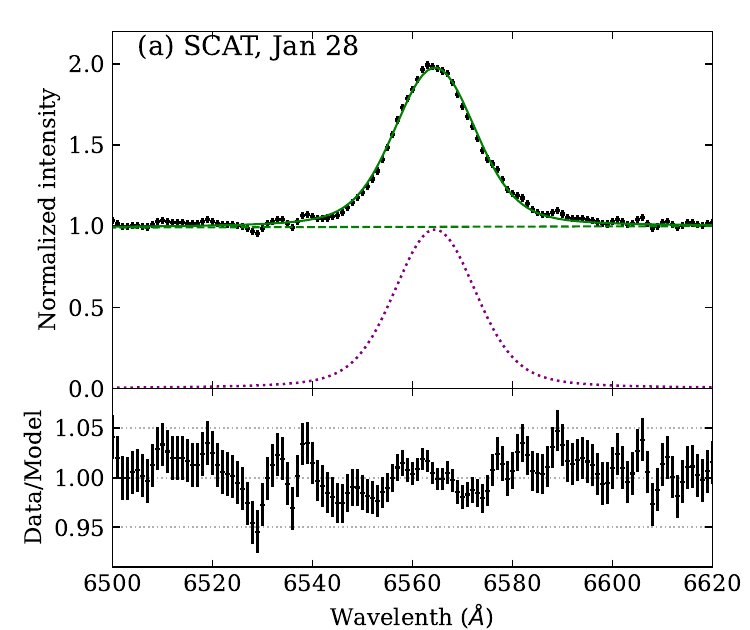}
\includegraphics[width=60mm]{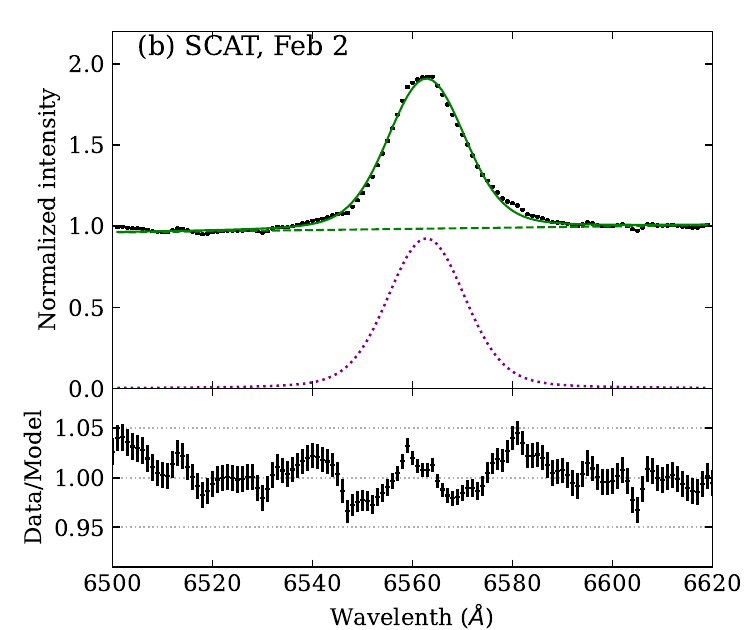}
\includegraphics[width=60mm]{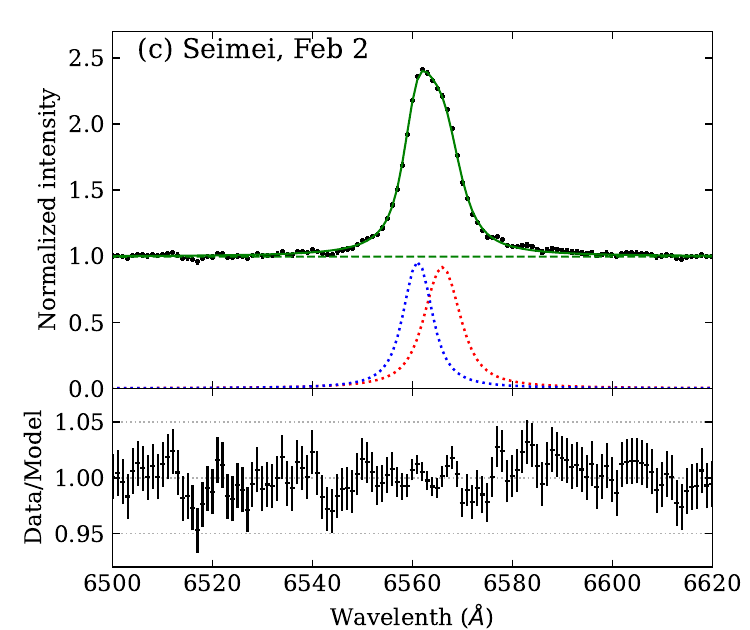}
\includegraphics[width=60mm]{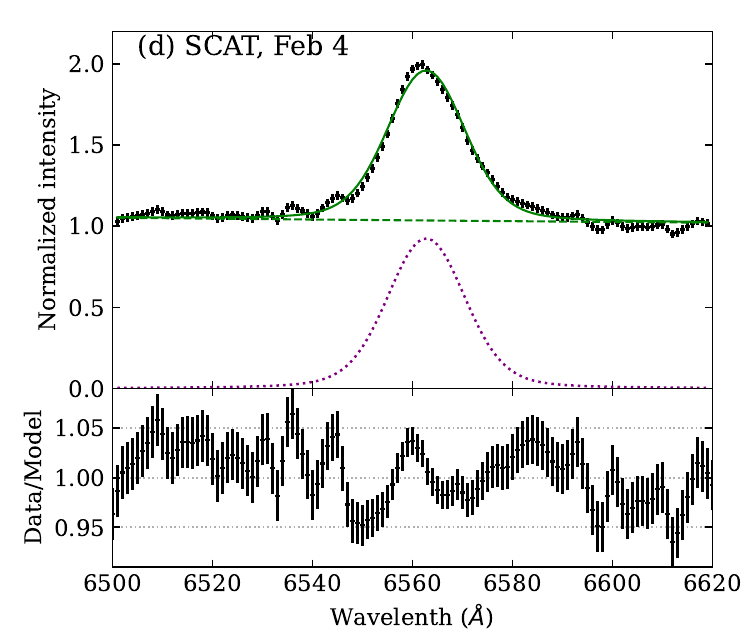}
\includegraphics[width=60mm]{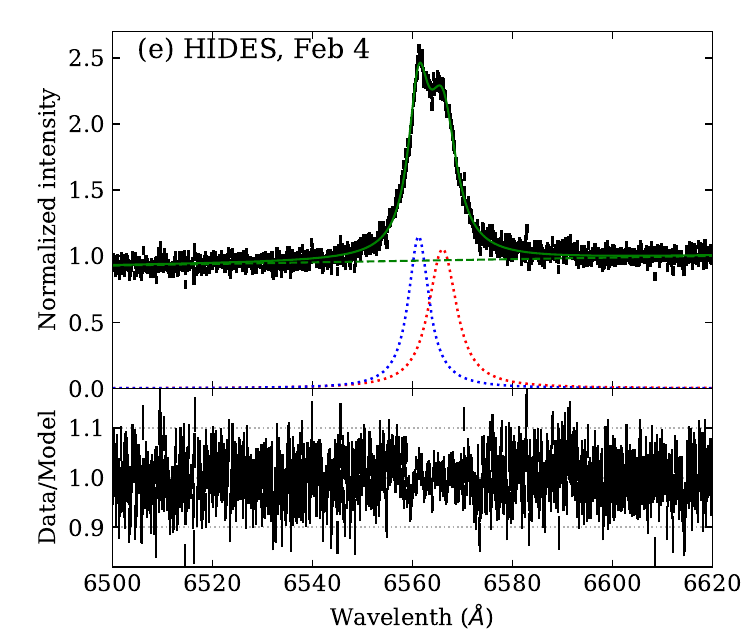}
\includegraphics[width=60mm]{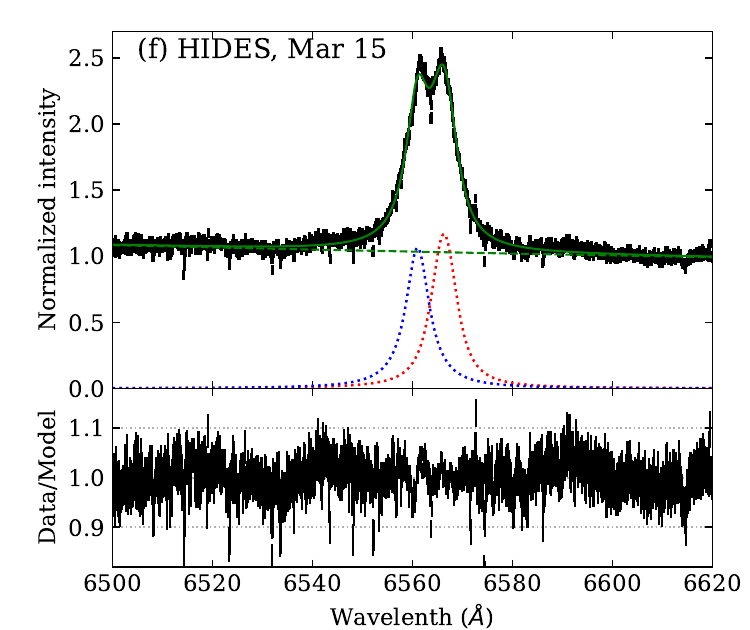}
\caption{H$\alpha$ line profiles obtained with the SCAT, Seimei/KOOLS-IFU and OAO/HIDES at different epochs and their best-fit models folded with the instrumental responses. For comparison of the results from different observatories at the same nights, the SCAT and Seimei/KOOLS-IFU data on February 2 and SCAT and OAO/HIDES on February 4 are presented. 
The contributions of the continuum and the Voigt components and the continuum component are shown separately with dotted and dashed lines, respectively. Bottom panels show the data versus model ratios.
{Alt text: Plots of Halpha line data and their best-fit models. The figure is separated into 6 panels labeled from (a) to (f), each of which presents the spectrum and model obtained with an instrument at one epoch. Each panel has a sub-panel at the bottom, plotting the data versus model ratio.}
\label{fig:halpha_fit}}
\end{figure*}

We performed spectral fitting of the observed H$\alpha$ line profiles.
We first applied a single Gaussian model to the SCAT data 
and a double Gaussian model to the Seimei/KOOLS-IFU 
and OAO/HIDES data. A power-law model was adopted to reproduce 
the continuum component. We found that the line profiles 
were roughly characterized with these models. However, as 
shown in the middle panels in Fig.~\ref{fig:halpha_fit}, 
residual structures remained in both wings of the line 
in all spectra, suggesting the presence of a component 
with fast radial velocities that cannot be explained 
with a single or double Gaussian models. 

Then, we replaced the Gaussians to the Voigt functions,   
following previous works \citep[e.g.,][]{Malacaria2017,Chhotaray2023}. 
These single or double Voigt function 
models resulted in better fits of the line wings in all spectra 
and allowed us to obtain the parameters to describe the overall 
line shapes and their variations. The peak-to-peak 
separation obtained from the Seimei/KOOLS-IFU and OAO/HIDES 
was typically $\sim 5$~\AA~, which corresponds to $\sim 230$ 
km s$^{-1}$. 
We also calculated the half width at zero intensity (HWZI)
of the entire H$\alpha$ line. Considering the quality of the spectra (Fig.~\ref{fig:halpha_fit}), we defined the wavelength region where the 
line is significantly detected as the range where the intensity of 
the best-fit H$\alpha$ line model (including both the two Voigt 
components for the OAO/HIDES and the Seimei/KOOLS-IFU) in which the instrumental responses are removed, exceeds 5\% of the continuum intensity, and estimated the HWZI as a half of this wavelength range. The typical HWZI was found to be $\sim 20$ \AA, which is converted to a velocity dispersion of $\sim 900$ 
km s$^{-1}$. Representative spectra of each observatory and their 
best-fit models are shown in Figure~\ref{fig:halpha_fit}
and the best-fit parameters are listed in 
Table~\ref{tab:fit_halpha}. 

To check the consistency between the different instruments, 
we compared the data and best-fit models taken on the same dates: 
February 2 and February 4 (see Appendix). We found that they were 
largely in agreement with each other at least on these two days. 
The SCAT spectra in the later period, however, showed a large 
scatter in the line width (or the HWZI) and gave somewhat smaller 
values compared with those from the OAO/HIDES and Seimei/KOOLS-IFU 
data obtained on the dates close to the SCAT observations, 
as seen in Table~\ref{tab:fit_halpha}. 
Considering this discrepancy and the uncertainty in the wavelength 
of the SCAT, we decided to use the SCAT results only for discussing 
the variation of the equivalent width (EW) of the line, which 
would be less affected by the instrumental response, 
and refer to the OAO/HIDES and Seimei/KOOLS-IFU results for the other 
spectral parameters.

\begin{figure}
\includegraphics[width=85mm]{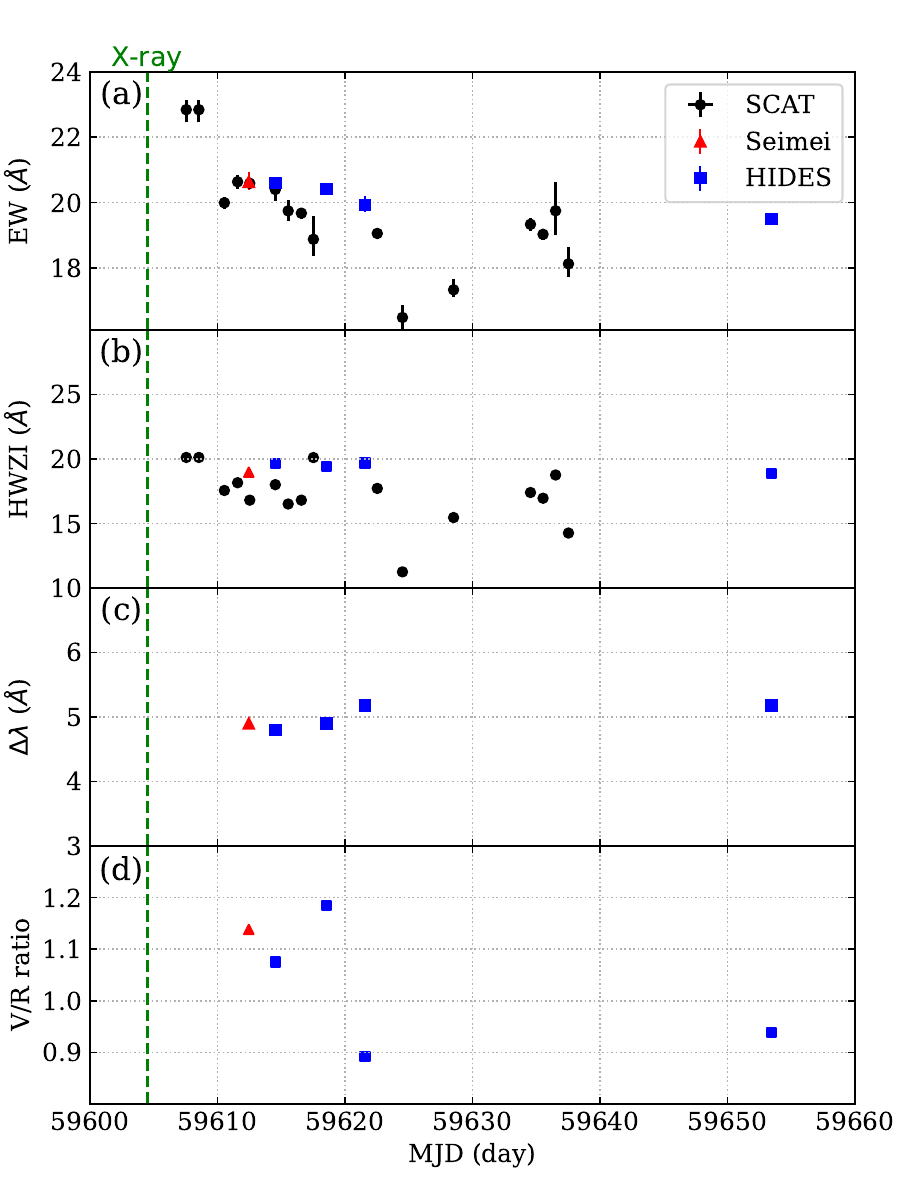}
\caption{Evolution of characteristic parameters: the equivalent width (a) and the HWZI (b) of the entire line structure, and the peak separation (c) and the $V/R$ ratio (d) of the double peak component. The V/R ratios were estimated from the intensities of the best-fit model (see text). The positive values indicate emission in the panel (a).
{Alt text: Graphs showing the time variation of spectral parameters, with 4 panels vertically lined-up and labeled from (a) to (d), which show different parameters.}
\label{fig:halpha_ew}}
\end{figure}

Figure~\ref{fig:halpha_ew}(a) shows the temporal 
evolution of the EW. The EW shows 
a clear trend of decline for $\sim 3$ weeks after 
the X-ray flaring event, with an amplitude of $\sim 5$ \AA. 
Then it slightly increased and finally became almost constant. 
Figure~\ref{fig:halpha_ew}(b) and (c) show the HWZI and the 
peak separation of the double peak component, respectively. 
No clear long-term increase or decrease trends are seen  
in these parameters.  
In Fig.~\ref{fig:halpha_ew}(d), 
we plot the trend in the V/R ratio, which is 
the intensity ratio at violet (V) and red (R) peaks, 
estimated from the best-fit model of 
the Seimei/KOOLS-IFU and OAO/HIDES data. In this 
estimation, we adopted the intensities of the line 
component (the sum of the two Voigt components, normalized 
by the continuum intensity) at the peak of the two Voigt models. 
Although the number of the data points are limited, 
the $V/R$ ratio decreased and then became constant,
similarly to the trend in the EW, confirming the 
association of the EW with the decline of the blue peak.

\subsection{Flux Variations observed with TESS}
\begin{figure*}
\includegraphics[width=170mm]{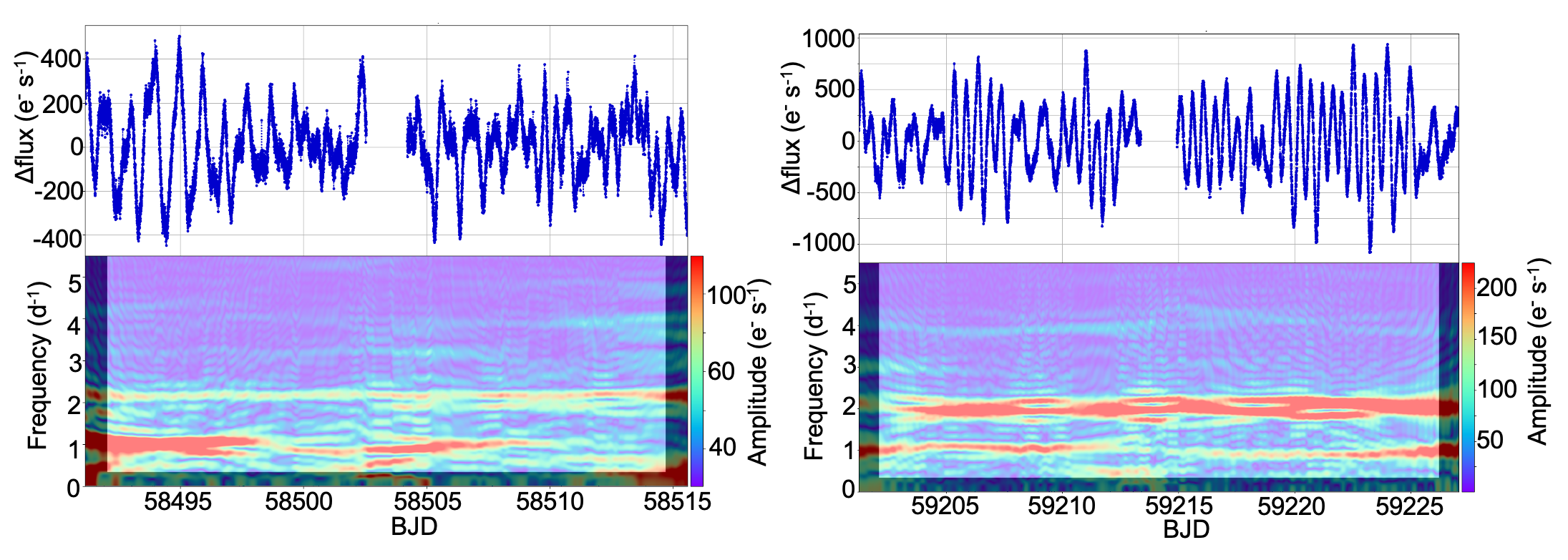}
\caption{TESS timing data in Epoch-1 (left) and Epoch-2 (right). Top and bottom panels show the light curves detrended with the running average over a 3-day window (see Section~\ref{sec:obs}) and their dynamic power spectra, respectively. 
{Alt text: Graphs of light curves and dynamic power spectra at two epochs. The figure is separated into two subfigures horizontally lined up, which show the results at different epochs. The subfigures are composed of two panels stacked vertically, the top and bottom of which present the light curves and the dynamic power spectra, respectively. The power spectra are shown as color contours, with the BJD in horizontal axis and the frequency in the vertical axis.}
\label{fig:tess_lc}}
\end{figure*}

\begin{figure}
\includegraphics[width=82mm]{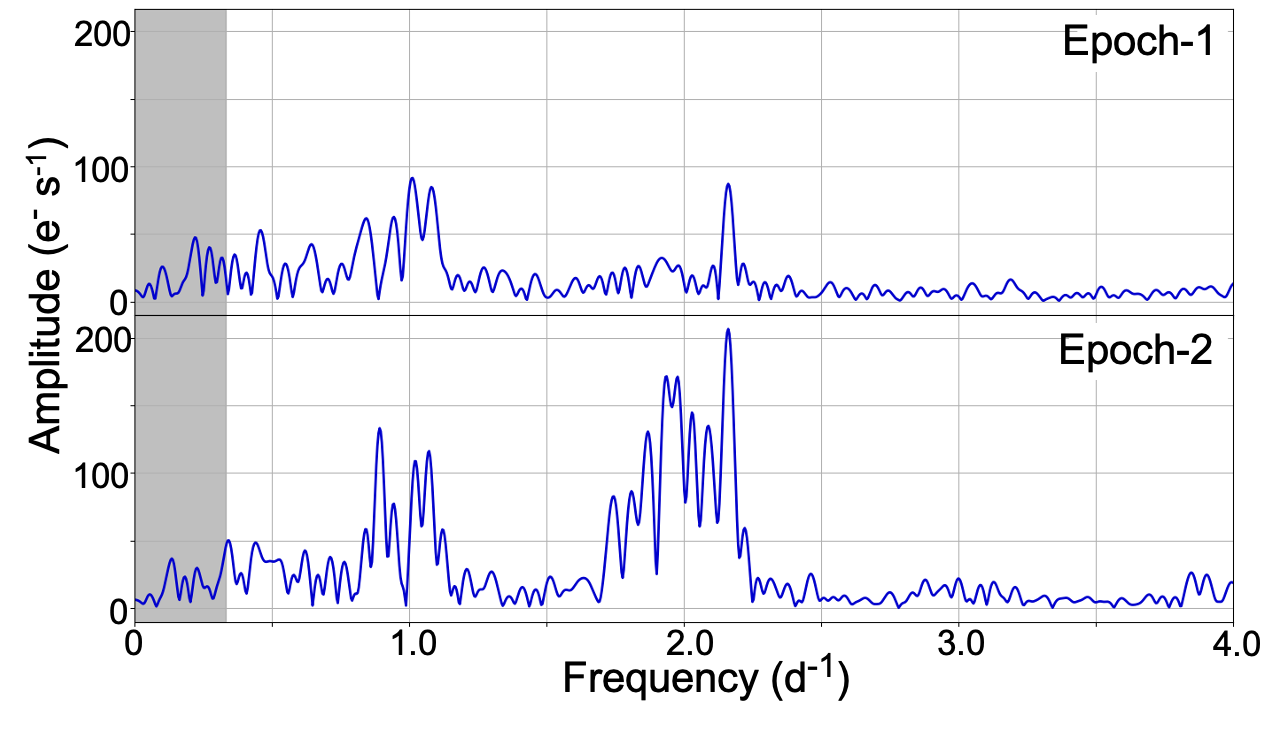}
\caption{Periodograms obtained from the TESS data (Fig.~\ref{fig:tess_lc}) in Epoch-1 (top) and Epoch-2 (bottom). The grey region ($< 1/3$ d$^{-1}$) shows the frequency range affected by the 3-day-window detrending. 
{Alt text: Figure showing periodograms, with two vertically-stacked panels presenting the data at the individual epochs.}
\label{fig:tess_psd}}
\end{figure}

In Figure~\ref{fig:tess_lc} we plot the light curves and dynamic power spectra obtained with TESS in Epoch-1 and Epoch-2. The dynamic power spectra were made with 30-minute bins using the light curve data within 5 days. 
In both epochs, the source exhibited strong variations on timescales of $\lesssim 1$ day, which were not strictly periodic and were accompanied by amplitude changes. Figure~\ref{fig:tess_psd} shows the periodograms made from the entire periods of the individual epochs. Both periodograms have peaks at frequencies of $\sim 1$ d$^{-1}$ and $\sim 2$ d$^{-1}$. As shown in the bottom panels of Fig.~\ref{fig:tess_lc}, the variations with these two typical frequencies are seen over almost the entire observations. In addition to these strong peaks, we found weak broad components in the periodograms, especially at low frequencies below $\sim 1-2$ d$^{-1}$.

\begin{table*}
\begin{threeparttable}
    \centering
    \tbl{The best-fit parameters of the H$\alpha$ line}{
    \begin{tabularx}{145mm}{cc ccccc}
    \hline \hline
%\multirow{2}{*}{Instrument (Date)} 
Instrument, Date & Component 
& $\lambda_{\rm cen}$ (\AA) % \multirow{1}{*}{$\lambda_{\mathrm cen}$}
& $\sigma$ (\AA) % \multirow{1}{*}{$\lambda_{\mathrm cen}$}
& $\gamma$ (\AA)
& Equivalent Width (\AA)
& HWZI\tnote{$\dagger$} (\AA)
%& $\chi^2/$d.o.f 
\\ \hline
\multicolumn{7}{l}{SCAT\tnote{*}} \\ 
Jan 28 & -  & $6564.5^{+0.1}_{-0.2}$ & $<0.9$ & $5.8^{+0.2}_{-0.3}$ & $22.8^{+0.3}_{-0.4}$ & $20.1$ \\
Jan 29 & -  & $6564.5^{+0.1}_{-0.2}$ & $<0.9$ & $5.8^{+0.2}_{-0.3}$ & $22.8^{+0.3}_{-0.4}$ & $20.1$ \\
Jan 31 & -  & $6561.7 \pm 0.1$ & $<0.6$ & $5.0 \pm 0.2$ & $20.0 \pm 0.2$ & $17.6$ \\
Feb 01 & -  & $6561.0^{+0.2}_{-0.1}$ & $<0.7$ & $5.2 \pm 0.2$ & $20.6 \pm 0.2$ & $18.2 $ \\
Feb 02 & -  & $6562.8 \pm 0.1$ & $<1.2$ & $4.4^{+0.1}_{-0.2}$ & $20.6 \pm 0.2$ & $16.8 $ \\
Feb 04 & -  & $6562.8^{+0.2}_{-0.1}$ & $<0.8$ & $5.1^{+0.2}_{-0.4}$ & $20.4 \pm 0.3$ & $18.0 $ \\
Feb 05 & -  & $6562.9^{+0.1}_{-0.2}$ & $<0.9$ & $4.5 \pm 0.3$ & $19.7 \pm 0.3$ & $16.5 $ \\
Feb 06 & -  & $6564.4^{+0.4}_{-0.1}$ & $<0.5$ & $4.7^{+0.1}_{-0.2}$ & $19.7 \pm 0.2$ & $16.8 $ \\
Feb 07 & -  & $6563.3^{+0.4}_{-0.2}$ & $<1.6$ & $7.0^{+0.6}_{-0.7}$ & $18.9^{+0.7}_{-0.5}$ & $20.1 $ \\
Feb 12 & -  & $6565.0 \pm 0.1$ & $<0.7$ & $5.4 \pm 0.2$ & $19.1 \pm 0.2$ & $17.7 $ \\
Feb 14 & -  & $6560.4 \pm 0.2$ & $2.8^{+0.4}_{-0.5}$ & $2.0 \pm 0.4$ & $16.5 \pm 0.4$ & $11.3 $ \\
Feb 18 & -  & $6562.8^{+0.2}_{-0.1}$ & $<0.8$ & $4.4^{+0.2}_{-0.4}$ & $17.3^{+0.3}_{-0.2}$ & $15.5 $ \\
Feb 24 & -  & $6560.8 \pm 0.1$ & $<0.7$ & $5.1 \pm 0.2$ & $19.3 \pm 0.2$ & $17.4 $ \\
Feb 25 & -  & $6563.0^{+0.1}_{-0.3}$ & $<0.6$ & $4.9^{+0.1}_{-0.2}$ & $19.0^{+0.1}_{-0.2}$ & $17.0 $ \\
Feb 26 & -  & $6563.5^{+0.3}_{-0.2}$ & $<1.6$ & $5.5^{+0.7}_{-0.6}$ & $19.8^{+0.9}_{-0.8}$ & $18.8 $ \\
Feb 27 & -  & $6561.7 \pm 0.2$ & $<1.1$ & $3.6^{+0.4}_{-0.5}$ & $18.1^{+0.5}_{-0.4}$ & $14.3 $ \\
\multicolumn{7}{l}{Seimei/KOOLS-IFU\tnote{*}} \\ 
Feb 2 & red & $6566.0 \pm 0.1$ & $<0.6$ & $6.7^{+0.2}_{-0.4}$ & $11.6 \pm 0.2$ & \multirow{2}{*}{$19.0 $} \\
 & blue & $6561.1^{+0.0}_{-0.1}$ & $<0.5$ & $4.6^{+0.1}_{-0.2}$ & $9.0^{+0.2}_{-0.1}$ &  \\
\multicolumn{7}{l}{OAO/HIDES\tnote{*}} \\ 
Feb 4 & red & $6566.02^{+0.04}_{-0.02}$ & $0.4 \pm 0.2$ & $6.46^{+0.08}_{-0.05}$ & $11.3 \pm 0.1$ & \multirow{2}{*}{$19.7 $} \\
 & blue & $6561.23^{+0.03}_{-0.02}$ & $< 0.2$ & $4.98^{+0.06}_{-0.04}$ & $9.3 \pm 0.1$ &  \\
Feb 8 & red & $6566.13^{+0.02}_{-0.05}$ & $1.57^{+0.06}_{-0.04}$ & $4.95 \pm 0.07$ & $9.9 \pm 0.1$ & \multirow{2}{*}{$19.4 $} \\
 & blue & $6561.23^{+0.03}_{-0.02}$ & $<0.2 $ & $5.68^{+0.05}_{-0.06}$ & $10.6 \pm 0.1$ &  \\
Feb 11 & red & $6566.27 \pm 0.03$ & $ <0.3 $ & $6.02^{+0.07}_{-0.10}$ & $10.6^{+0.3}_{-0.1}$ & \multirow{2}{*}{$19.7 $} \\
 & blue & $6561.09^{+0.03}_{-0.04}$ & $<0.2 $ & $5.88^{+0.08}_{-0.10}$ & $9.3^{+0.1}_{-0.2}$ &  \\
Mar 15 & red & $6566.37 \pm 0.03$ & $0.58^{+0.05}_{-0.16}$ & $5.33^{+0.05}_{-0.07}$ & $9.9 \pm 0.1$ & \multirow{2}{*}{$18.9 $} \\
 & blue & $6561.19 \pm 0.03$ & $<0.4$ & $5.76^{+0.10}_{-0.04}$ & $9.6 \pm 0.1$ &  \\
    \hline
    \end{tabularx}}
    \smallskip
     \begin{tablenotes}[normal]
      \footnotesize
      \small
    \item[*] % 
      Voigt function $+$ power-law model was adopted for the SCAT, and Voigt function (red) $+$ Voigt function (blue) $+$ power-law model were adopted for the Seimei/KOOLS-IFU and OAO/HIDES.
    \item[$\dagger$] % 
      Half width at zero intensity (HWZI) of the entire H$\alpha$ line, calculated as a half of the wavelength range where the intensity of the best-fit emission line model (including both the blue and red components for Seimei/KOOLS-IFU and OAO/HIDES) exceeds 5\% of the continuum intensity.   
    \end{tablenotes}
    \label{tab:fit_halpha}
\end{threeparttable}
\end{table*}

\section{Discussion} 
\label{sec:discussion}
\subsection{Origin of the H$\alpha$ emission line}
We have conducted an optical spectroscopic monitoring of the 
X-ray binary \srcname~(LY CMa) for about 1.5 months after 
the X-ray flaring event. The observed spectrum always 
showed a strong H$\alpha$ emission line as 
reported by \citet{bhattacharyya2022} and 
\citet{sugizaki2022}, in which the existence of a Be 
circumstellar disk was suggested. Indeed, our high 
resolution spectroscopy revealed a double-peaked profile, 
which is often seen in Be stars \citep[e.g.,][]{Struve1931,Hanushik1996,Porter2003}
and is generally explained by a nearly Keplerian disk 
formed around the star. 
We therefore consider that the observed H$\alpha$ line is mainly 
produced by the Be disk. 

From the observed peak-to-peak 
separation ($\sim$ 5~\AA), the  
projected rotational velocity of the Be disk 
($v \sin i$, where $v$ is the rotational velocity 
and $i$ is the inclination angle) is estimated 
to be $v \sin i \sim 110$ km s$^{-1}$, following 
Equation (14) in \citet{Dachs1986}. 
Assuming the star is an evolved B-type star with 
a radius of $R \sim 11.8 R_\odot$ and a surface 
gravity of $\log g \sim 3$, which were obtained 
by \cite{bhattacharyya2022} through SED fitting, 
we estimated its critical velocity of the rotation 
to be $\sim 290$ km s$^{-1}$. The $v \sin i$ value estimated 
from the H$\alpha$ line does not exceed this value and 
hence is consistent with the rotation velocity of the 
Be disk, unless $i$ is smaller than $\sim 20^\circ$. 
Note, however, that a stellar rotation velocity of $v_\mathrm{rot} \sin i \sim 93$ km s$^{-1}$ was obtained in the APOGEE Data Release 16 \citet{Jonsson2020}. This value, although the information of its uncertainty was not provided, is somewhat smaller than the velocity estimated from the H$\alpha$ line. If this is the case, the Be disk may be warped around the region emitting the $H\alpha$ line and have a larger inclination angle than the spin axis of the star. 
In our analysis, we adopted the Voigt function(s) 
to model the line profile. This is only an empirical model, 
and application of a physical model is required 
to test if the line profile is really explained by the Be 
disk and discuss the Be disk structure, which is beyond the 
scope of this paper and we leave it as a future work.
We also do not rule out that 
the H$\alpha$ line emission is contributed to 
some extent by other structures such as a stellar wind 
and the accretion disk of the compact object (see below). 

In addition to the main double-peaked component, we 
detected a broad component extending its both sides. 
Modelling the entire line profile with a single 
or double Voigt function, we obtained 
a HWZI of $\gtrsim 900$ km s$^{-1}$. Similar broad 
wings were observed in BeXRBs 4U 0115$+$63 
\citep{Negueruela2001, Reig2007}, A 0538$-$66 
\citep{Rajoelimanana2017}, and IGR J06074$+$2205 
\citep{Chhotaray2024}.
In some of the works, the broad wings were explained by 
warping of the Be disk; the inner regions of the 
Be disk has a larger inclination angle than  
the outer regions and thus have a high line-of-sight 
velocity, which results in a broad line component. 
In our case, however, it is difficult to attribute  
to the rotational motion in the inner disk regions, 
because the observed velocity of the broad component 
is much larger than the critical velocity 
of the star ($\sim 290$ km s$^{-1}$)\footnote{We note that the 
stellar parameters used to estimate the critical 
velocity may have some uncertainty 
caused by e.g., contribution of the Be disk emission, 
which is ignored in the SED fitting, and uncertainty in interstellar extinction. 
Nevertheless we assume $v \sin i 
\sim 300$ km s$^{-1}$ and the stellar parameters 
from \cite{bhattacharyya2022} in the following 
discussions.} and the value reported in the
APOGEE catalog ($\sim 93$ km s$^{-1}$; \citealt{Jonsson2020}).

One possibility of its origin would be a stellar 
wind. The terminal velocity $v_{\infty}$ of a 
radiatively driven wind is roughly $\sim$ 2 times 
the escape velocity \citep[e.g.,][]{Abbott1978,Crowther2006}. 
For \srcname, the escape velocity is $\sim 400$ km s$^{-1}$ 
assuming the stellar parameters in \citet{bhattacharyya2022} so  
$v_{\infty}$ is estimated to be 
$\sim 800$ km s$^{-1}$. This value is comparable 
to the velocity that we estimated from the H$\alpha$ line.
Hence, the stellar wind could explain the observed 
broad component more naturally than the disk warping. 
The detection of rapid X-ray spectral variability 
like SFXTs (\citealt{sugizaki2022}; Sugizaki et al. in prep) 
is in agreement with the idea that a strong stellar wind 
is present in this system. We note, however, that no 
observed lines show a blueshifted absorption component 
or the p-Cygni profile, which are often seen in the lines 
produced by stellar winds. 

Another possibility is the accretion disk of the 
compact object. Optical emission lines with similar 
widths are frequently detected in low mass X-ray binaries
\citep[e.g.,][]{Casares1995,Jimenez-Ibarra2019}, 
even at an X-ray luminosity as low as $\sim 10^{33}$ 
erg s$^{-1}$ \citep{yoshitake2022}. They are  
considered to be produced in the outer region of 
the accretion disk, irradiated by X-rays from the 
inner disk region. Although lines from the accretion 
disk are not usually visible in HMXBs, 
there is one clear evidence from the Be/black hole 
binary MWC 656 \citep{casares2014}, where 
the HeII ($\lambda4648$) line exhibited the 
radial velocity modulation by the orbital motion of the accretion 
disk. In our observations of~\srcname, we did not detect such 
periodic radial velocity modulation, but this is likely 
due to the limited number of the high resolution spectra 
and insufficient orbital phase coverage. We also 
found that the HeII line was not significantly detected 
in any of the spectra in~\srcname, unlike MWC 656.  
This difference could be explained if they have 
different X-ray spectral profiles irradiating the 
outer, optical emitting region of the disk, which 
could result in different ionization degrees. Another 
possibility is that the disk in \srcname~was truncated 
outside the HeII emitting region due to the magnetic 
field of the compact object.

If the broad H$\alpha$ component really originates
in the accretion disk, its maximum velocity ($\sim$ 
1000 km s$^{-1}$) corresponds to the Keplerian 
velocity at a radius $R_{\rm K} \sim 2 \times 10^{10} 
\left(M/1.4 M_\odot \right)$ cm, where $M$ 
is the mass of the compact object. 
Many of the known BeXRBs and sgXBs host a magnetized 
neutron star. If it is also the case in \srcname, 
the accretion disk should be truncated of the order of 
the Alfv\'en radius $R_{\rm A}^{(0)}$, which is the 
radius where the pressure of the accreted gas balances 
the magnetic pressure: 
\begin{equation}
R_{\rm A}^{(0)} = \mu^{4/7} (2GM)^{-1/7} \dot{M}^{-2/7} 
\end{equation}
where $\mu$, $G$, and $\dot{M}$ represent the magnetic dipole 
moment of the neutron star, the gravitational constant, 
and the mass accretion rate, respectively \citep{Elsner1977}. 
This is obtained under spherical accretion conditions  
and the actual inner radius of the accretion disk 
is $R_{\rm in} \sim 0.5 R_{\rm A}^{(0)}$ \citep{Ghosh1979a,Ghosh1979b}. 
Substituting $\mu = BR_{\rm NS}^3$ and 
$\dot{M} = LR_{\rm NS}/(GM)$ (where $B$ and 
$R_{\rm NS}$ are the magnetic field strength 
and the radius of the neutron star, and $L$ is 
the luminosity) and adopting the typical value 
of $B = 10^{12}$ G and an X-ray luminosity of 
$\sim 10^{32}$ erg s$^{-1}$, which is the lowest 
value obtained by \citet{sugizaki2022} 
after the X-ray flaring event, we obtain 
\begin{align}
R_{\rm in} \sim & 10^{10} \left(\frac{M}{1.4 M_\odot}\right)^{1/7} \left(\frac{R}{10^6~{\rm km}}\right)^{10/7}  \nonumber \\
& \left(\frac{B}{10^{12}~{\rm G}}\right)^{4/7} \left(\frac{L}{10^{32}~{\rm erg~s}^{-1}}\right)^{-2/7} \,{\rm cm}, 
\end{align}
which is consistent with $R_{\rm K}$. 
In addition, the inner disk radius should be smaller 
than the co-rotation radius: 
\begin{equation}
R_{\rm C} = \left( \frac{GMP^2}{4 \pi^2} \right)^{1/3} \sim 
 10^{10} \left(\frac{M}{1.4 M_\odot}\right)^{1/3} 
\left(\frac{P}{1000~{\rm s}} \right)^{2/3}~{\rm cm}.
\end{equation}
where $P$ is the spin period of the neutron star;  
otherwise the accretion disk is evaporated due to 
the propeller effect \citep{Illarionov1975}. 
The condition $R_{\rm C} > R_{K}$ 
can be realized with $P \gtrsim 10^3$ s, and 
some of BeXRBs and SFXTs actually have such 
relatively long spin periods 
\citep[e.g.,][]{chaty2011}. 
Considering the above estimations, we conclude that 
the accretion disk interpretation is possible even if 
the compact star is a magnetized neutron star.

\subsection{evolution of the H$\alpha$ line}

The EW of the H$\alpha$ line was found to decrease 
in $\sim 3$ weeks from the X-ray activity, 
then slightly increased, and finally became almost constant.  
The OAO/HIDES data indicate that the decrease was 
associated with the blue peak of the double-peaked 
component, rather than the red peak and the broad 
component, which were almost constant over the 
entire period of the observation campaign.
Meanwhile, the V and Ic-band magnitudes 
were almost constant, and therefore the 
continuum spectrum is unlikely to have 
varied significantly. This suggests that the 
structure of the Be disk did not change largely in this 
period, and the variation of the blue peak component 
should be produced by a local change in the structure of 
the Be disk or variation of other structures such as 
the stellar wind. 

A possible interpretation is that 
a small perturbation of the Be disk induced by the 
passage of the compact object. If such a 
perturbation occurred in the approaching part of 
the disk, the blueshifted component of the H$\alpha$ 
line could be enhanced and then gradually weakened 
with the relaxation of the perturbation. 
A similar behaviour in the H$\alpha$ line profile 
was observed in the Be X-ray binary A 0535$+$262 
\citep{Moritani2011}, 
where the observed change in the ``blue shoulder'' 
component after the periastron passage was 
attributed to the gas stream from the Be disk to the neutron star. 
An alternative interpretation is that dense clumps 
of a stellar wind 
was approaching towards us, which contributed 
a fraction of the blue peak component in addition to 
the emission from the Be disk and caused its 
enhancement. In this case, the decrease in the blue 
peak could be explained by the decrease in the 
density of the clumps due to their expansion. 
In either of the two possibilities, it is not 
easy to explain the slight increase of the EW 
after the decline, but a small perturbation 
might have occurred in the structures that 
caused the enhancement of the EW.

It is natural to consider that the observed EW decrease was 
triggered by the X-ray flare (i.e., the episode of 
accretion onto the compact object), given the fact that it 
occurred just after the X-ray flare and has a much shorter 
timescale than that of the magnitude variation, although 
different interpretations are not ruled out completely. 
In this sense, the turbulence of the Be disk may be more plausible scenario for the cause of the EW decrease, 
rather than the variation of other structures. Similar 
behavior is observed in some BeXRBs such as 4U 0115$+$63 
and A 0535$+$26 \citep{raig2011}, which showed H$\alpha$ 
EW decrease after giant outbursts. These sources tend to 
have a small binary size or a small orbital separation at periastron passage. If \srcname~is also a BeXRB 
accreting from the Be disk, it may share similar characteristics. Note, however, that \srcname~kept 
almost constant optical magnitudes during the variation 
H$\alpha$ EW, in contrast to 4U 0115$+$63 and A 0535$+$26, 
which showed significant magnitude decrease and larger drop of 
the H$\alpha$ EW \citep{raig2011, Haigh2004, Moritani2013}.

By contrast to the main double-peaked component, the broad 
wing did not vary significantly, over the entire period 
of our observations. This suggests that the structure that 
produced the broad component was kept unchanged for at 
least $\sim 1.5$ months after the X-ray flare, and therefore 
it should originate in the stable structure on this timescale.
If it was produced by the accretion disk as discussed above,  
low-level accretion continued for $\gtrsim 1.5$ months. 
Further optical spectroscopic and X-ray monitoring would be 
required to determine the origin and time evolution of 
the structure. 

\subsection{Origin of Optical Flux Variation}

A long-term optical flux variation was observed on timescales 
of a few thousand days (Fig.~\ref{fig:LC_longterm}). 
Its amplitude was larger in the redder 
band, $\sim 0.5$ mag in V band and $\sim 0.8$ mag in I band. 
Similar "redder-when-brighter" long-term variations on timescales 
of $\sim 10^3-10^4$ days have been observed in Be stars and 
is explained by the evolution of the Be disk \citep{Rajoelimanana2011}.
In~\srcname, \citet{bhattacharyya2022} suggested a possible correlation between the H$\alpha$ EW and the V-band magnitude 
on a timescale of $\gtrsim$ year (Fig.~5 in that work). 
We have confirmed it with the KWS long-term light curve, 
which provides the data after MJD $\sim$ 55000. 
This indicates that the Be disk is the main contributor 
to the H$\alpha$ emission line, although the 
accretion disk and/or the stellar wind could also 
contribute to it. 
The detection of X-ray activity and our optical 
spectroscopy correspond to the high flux phase in the 
long-term optical light curve, suggesting that the 
Be disk developed almost maximally in these epochs. 

Using TESS, we also detected flux variation 
on timescales of hours to a few days. 
The periodograms in 2019 and 2021 show 
groups of narrow peaks at frequencies of 
$\sim 1$ and $\sim 2$ day$^{-1}$, which have 
been detected in known BeXRBs \citep{Reig2022}
and isolated Be stars \citep{Labadie-Bartz2022}. 
These quasi-periodic variations are generally 
explained by the pulsation and/or rotation 
of the Be star. Assuming a critical velocity 
of $\sim$ 300 km s$^{-1}$ and the stellar radius 
of 11.8 $R_\odot$, we estimate the  
spin period of the Be star in~\srcname~to 
be $\sim 2$ days. This timescale is 
larger than the observed quasi-periodic 
variation, so stellar pulsation is more 
likely to be the origin of the variation. 

We also detected a broadened component 
seen particularly at low frequencies below 
$\sim 1-2$ d$^{-1}$. Previous observations 
detected a similar type of variations 
in early-type stars, which has been 
explained by inhomogeneities in the stellar 
surface or wind, or by internal gravity waves 
produced at the transition zone of the 
convective core and radiative envelope. 
In~\srcname, the strength of this variation 
is somewhat stronger at higher V and Ic-band 
fluxes, so the Be disk may also contribute 
to the variation. \citet{Reig2022} found that 
periodograms of sgXBs can be characterized by  
a red noise below $~2$ d$^{-1}$ and do not show  
strong peaked components, unlike BeXRBs. 
The TESS periodograms of \srcname~have 
similar properties to those of BeXRBs
rather than the sgXBs.

\section{Conclusion}
Our optical spectroscopic monitoring of the X-ray 
transient \srcname~(LY CMa), conducted for 1.5 months 
after the X-ray activity was detected, has provided 
information to understand the circumstellar structure 
and its variation in the binary system. The observed 
H$\alpha$ line has a double-peaked profile, which is 
likely to originate mainly in the Be disk. The blue 
peak was found to decrease in the first $\sim$ 3 weeks.  
We interpreted its origin as a turbulence in the Be disk 
caused by the compact object. In addition to the double-peaked 
component, a broad wing was detected in the entire observation 
period. Its maximum velocity exceeds the break-up velocity 
of the Be star, and hence the component was likely 
to be produced by structures such as a stellar wind from 
the donor star or the accretion disk, rather than the Be disk. 
There are, however, other interpretations that are not completely ruled out, for the H$\alpha$ line profile and its variation. Further spectroscopic monitoring is needed  
to fully uncover the circumstellar structure around the 
high-mass donor star and determine the true nature 
of the binary system. We also studied long-term ($\gg$ 1 day) 
and short-term ($\sim$ 1 day) optical flux variations and 
found they have similar properties to other BeXRBs.

\begin{ack}
We thank Atsuo Okazaki for discussion of stellar winds.
This research has made use of software provided by 
the High Energy Astrophysics Science Archive Research 
Center (HEASARC), which is a service of the Astrophysics Science Division at NASA/GSFC. 
This paper includes data collected with the TESS mission, obtained from the MAST data archive at the Space Telescope Science Institute (STScI). Funding for the TESS mission is provided by the NASA Explorer Program. STScI is operated by the Association of Universities for Research in Astronomy, Inc., under NASA contract NAS 5–26555.  
Part of this work was financially supported 
by Grants-in-Aid for Scientific Research 19K14762 (MS) 
from the Ministry of Education, Culture, Sports, 
Science and Technology (MEXT) of Japan. 
\end{ack}

%\section*{Funding}
% This research was supported by ...

%\section*{Data availability} 
% The data underlying this article are available ...  
% Sample Data Availability Statements 
% https://academic.oup.com/pages/open-research/research-data#Data%20Availability%20Statements

\bibliographystyle{apj}
\bibliography{maxij0709_ms}{}
% Any journal's BST file (e.g., apj.bst) can be used as PASJ's BST is unavailable.    
% \bibliographystyle{****}
% \bibliography{****}

\appendix %%%%%%%%%%%%%%%%%%%%%%%%%%%%%%%%%%%%%%%%%%%%%%%%%%%%%%%%
\section*{Comparison of Spectral-Fitting Results Between Different Instruments}

\begin{figure}[thb]
\begin{center}
\includegraphics[width=85mm]{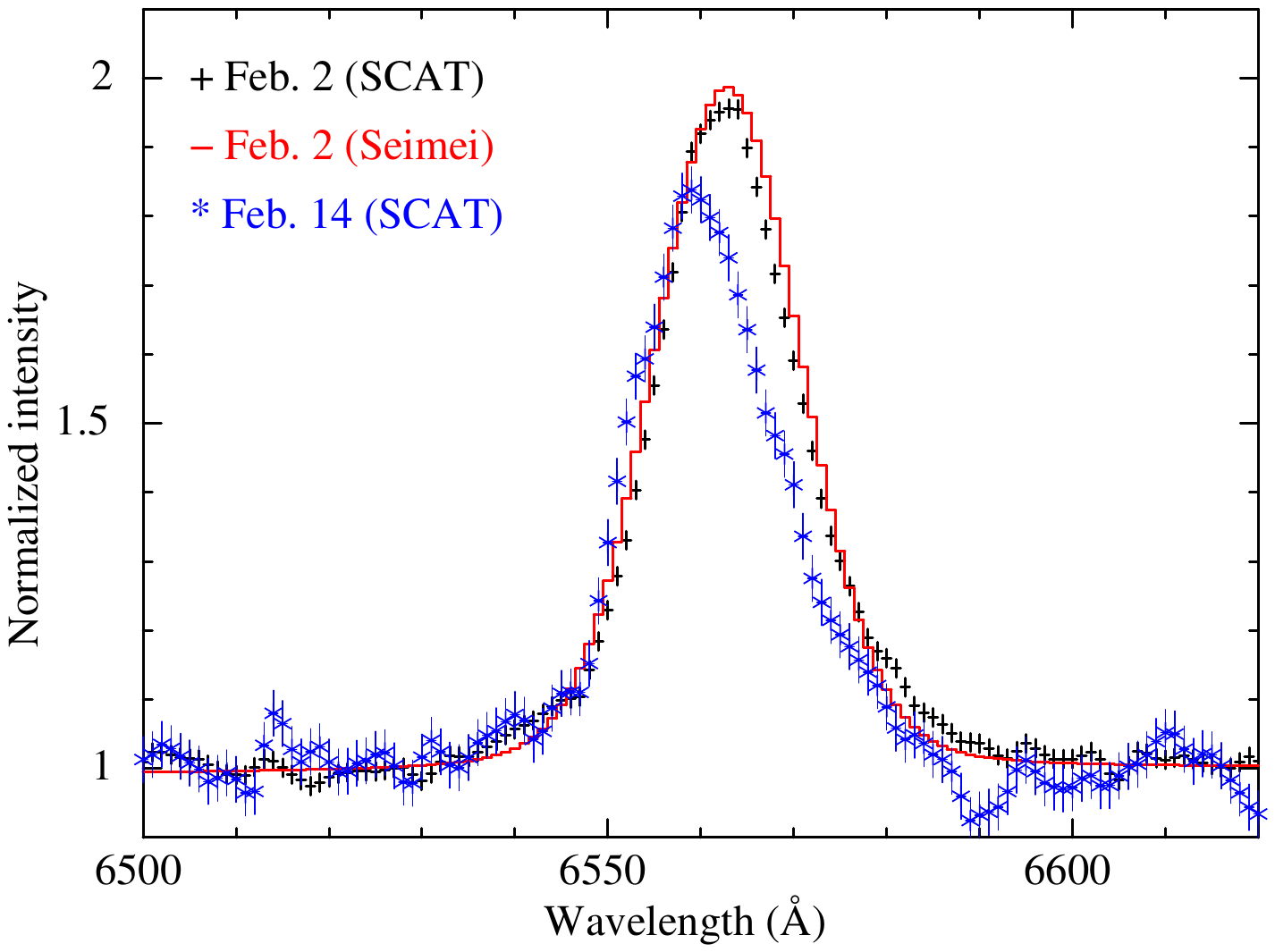}
\includegraphics[width=85mm]{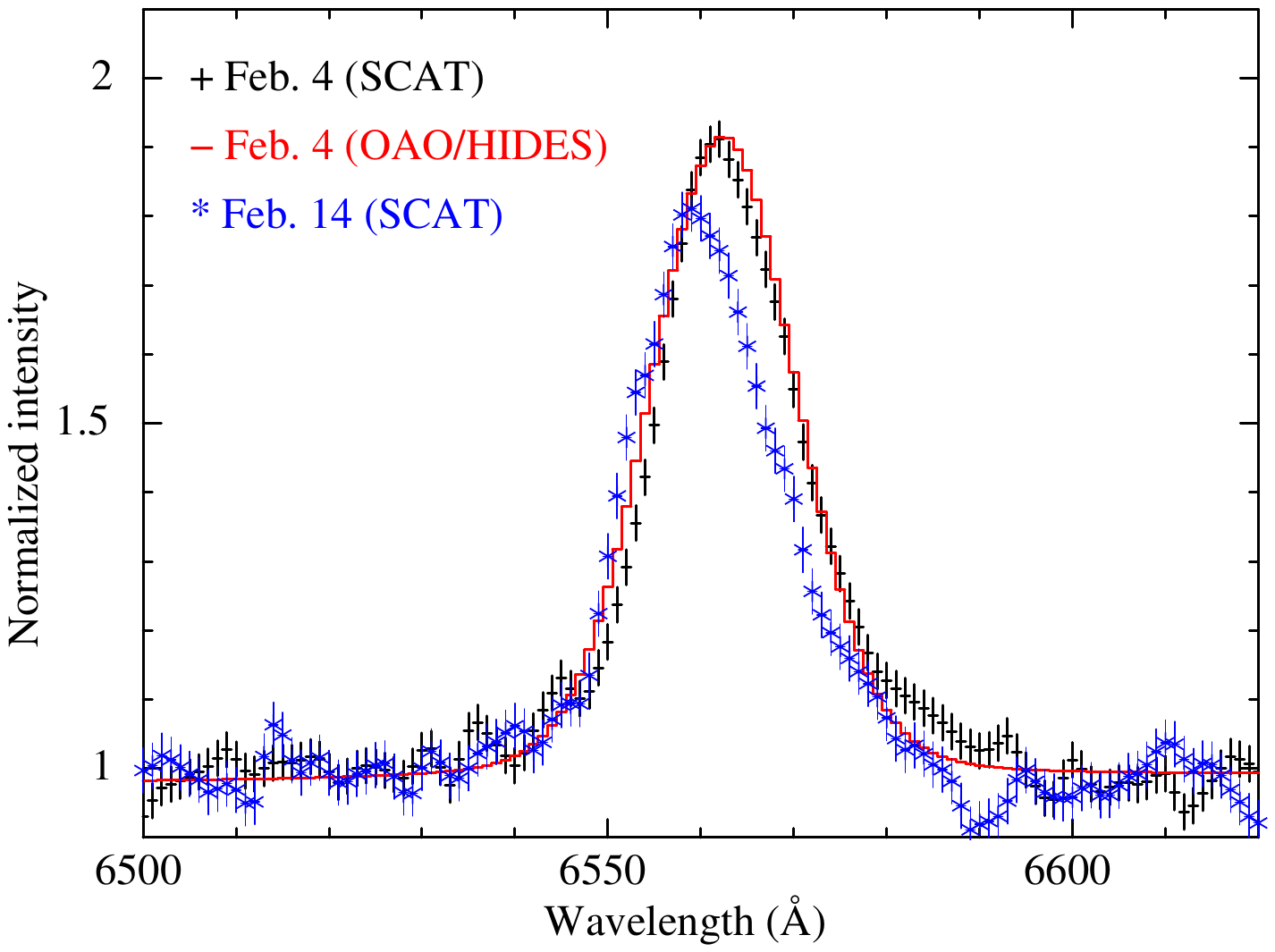}
 \end{center}
 \caption{
 SCAT spectra (black crosses) around the H$\alpha$ line obtained on February 2 (top) and February 4 (bottom) and the best-fit double Voigt model (red lines) obtained with the Seimei/KOOLS-IFU and the OAO/HIDES on the same dates, which are folded with the SCAT instrument response, respectively. The line center wavelengths of the two Voigt components are shifted by the same amount so that they better fits the SCAT profiles. The SCAT spectrum on February 14 (blue asterisks), which has the narrowest line profile, is also presented in both panels.
 {Alt text: Figure showing the profiles of an emission line located around 6600~\AA obtained with different instruments. It is composed of two panels, which present results at different nights.}
\label{fig:comp_fit}}
\end{figure}

Figure~\ref{fig:comp_fit} shows the SCAT data taken on February 2 and 4, compared with the best-fit double Voigt function models of the Seimei/KOOLS-IFU data and the OAO/HIDES data taken at the same nights, respectively. The instrumental responses of the SCAT (see Section~\ref{subsec:halpha}) are combined to the models.
To adjust the line center of the models to the SCAT data, which could have a significant uncertainty in the wavelength (Section~\ref{subsec:SCAT}), we shifted the line center wavelengths of the two Voigt components by $\Delta \lambda = 0.2$\AA~for February 2 and $-1.3$\AA~for February 4. 
The data are approximately in agreement with the model, while small discrepancies are seen around the peak and winds of the line. We also plotted the spectrum on February 14, which has the narrowest line profile. The variation of the line width in the SCAT data could be due to fluctuation of the focus of the spectrometer, although the possibility that the variation of the H$\alpha$ line itself could be contributed to some extend is not completely ruled out.

\begin{figure}[thb]
\begin{center}
\includegraphics[width=85mm]{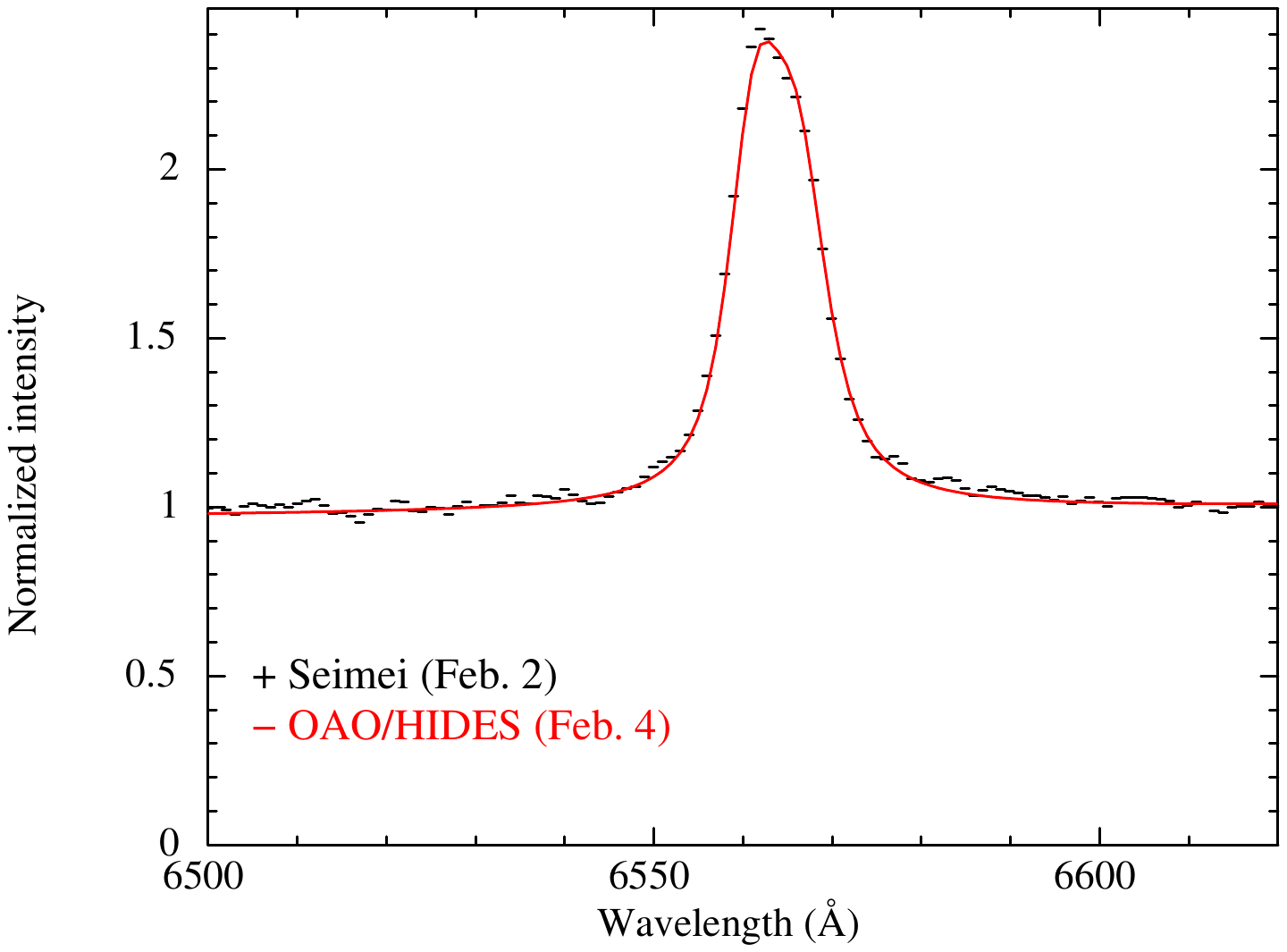}
 \end{center}
 \caption{Seimei/KOOLS-IFU spectrum on February 2 (black cross) compared with the best-fit model from the OAO/HIDES data on February 4 (red line) folded with the Seimei instrumental response.
  {Alt text: Plot of a spectrum and a model of the H$\alpha$ emission line.}
\label{fig:comp_fit_seimei}}
\end{figure}

In Figure~\ref{fig:comp_fit_seimei}, we compare the Seimei/KOOLS\_IFU data taken on February 2 and the best-fit model obtained with the OAO/HIDES spectrum on February 4. The Seimei spectrum is very well reproduced by the model, except around the peak, where the Seimei data have a small excess on the bluer side. This discrepancy could be explained by the variation of the source, because the decrease of the blue peak was detected in the OAO/HIDES data (Sec.~\ref{subsec:halpha}).

% No section number is necessary. Add ``*'' after \verb/\section/.

%%%% 
%\section{Case of two or more paragraphs}

% Text of appendix

\end{document}